\documentclass[aps,prb,twocolumn,groupedaddress,showpacs,floatfix]{revtex4-1}
\usepackage{graphicx,amsmath}
\usepackage{amsfonts}
\usepackage{amssymb}
\usepackage{nicefrac}

\begin{document}

\title{{\it Ab initio} study of  the two-dimensional
metallic state at the surface of SrTiO$_3$: importance of oxygen vacancies}
\author{Juan Shen}
\affiliation{Institut f\"ur Theoretische Physik, Goethe-Universit\"at Frankfurt, Max-von-Laue-Stra{\ss}e 1, 60438 Frankfurt am Main}
\author{Hunpyo Lee}
\affiliation{Institut f\"ur Theoretische Physik, Goethe-Universit\"at Frankfurt, Max-von-Laue-Stra{\ss}e 1, 60438 Frankfurt am Main}
\author{Roser Valent{\'\i}}
\affiliation{Institut f\"ur Theoretische Physik, Goethe-Universit\"at Frankfurt, Max-von-Laue-Stra{\ss}e 1, 60438 Frankfurt am Main}
\author{Harald  O. Jeschke}
\affiliation{Institut f\"ur Theoretische Physik, Goethe-Universit\"at Frankfurt, Max-von-Laue-Stra{\ss}e 1, 60438 Frankfurt am Main}
\date{\today}

\begin{abstract}
  Motivated by recent angle-resolved photoemission spectroscopy
  (ARPES) observations of a highly metallic two-dimensional electron
  gas (2DEG) at the (001) vacuum-cleaved surface of SrTiO$_3$ and the
  subsequent discussion on the possible role of oxygen vacancies for
  the appearance of such a state~\cite{Santander-2011-nature}, we
  analyze by means of density functional theory (DFT) the electronic
  structure of various oxygen-deficient SrTiO$_3$ surface slabs.  We
  find a significant surface reconstruction after introducing oxygen
  vacancies and we show that the charges resulting from
  surface-localized oxygen vacancies --independently of the oxygen
  concentration-- redistribute in the surface region and deplete
  rapidly within a few layers from the surface suggesting the
  formation of a 2DEG. We discuss the underlying model emerging from
  such observations.
\end{abstract}

\maketitle

\section{Introduction}

In recent years, SrTiO$_3$ (STO), which is a wide band-gap
semiconductor (3.2~eV), has been commonly used as a substrate for the
epitaxial growth of oxide heterostructures with other perovskite
compounds such as the band insulator LaAlO$_3$ (LAO) or the Mott
insulator LaVO$_3$. In 2004 Ohtomo and Hwang~\cite{Ohtomo-2004-nature}
detected a two-dimensional electron gas (2DEG) state at the interface
of a LAO/STO heterostructure. A polar catastrophe was suggested to be
the cause for this state.  Numerous experimental and theoretical
studies have been carried out on this unusual metallic state since
then~\cite{Ohtomo-2004-nature,Huijben-2006-nmat,Nakagawa-2005-nmat,Thiel-2006-science,Eckstein-2007-nmat,Reyren-2007-science,Popovic-2008-prl,Basletic-2008-nmat,Cen-2009-science,
  Janicka-2009-prl,Li-2011-science,Arras-2012-prb,Zhong_2012} and
fascinating features like superconductivity and coexistence with
ferromagnetism have been attributed to this 2DEG.  More recently,
Santander-Syro {\it et al.}~\cite{Santander-2011-nature} and Meevasana
{\it et al.}~\cite{Meevasana-2011-nmat} found that it is not necessary
to have an interface with a polar interface with other materials in
order to obtain a 2DEG at the STO surface. By simply fracturing bulk
STO and using angle-resolved photoemission spectroscopy, these
authors studied a vacuum cleaved STO (100) surface and found high
carrier densities at the bare surface.  While the authors' findings
opened a new scheme in oxide electronics, the precise nature of the
2DEG has remained elusive. Oxygen vacancies as n-type dopants and also
as a by-product of the cleaving process were suggested to be related
with the 2DEG~\cite{Santander-2011-nature}.  In fact, in a recent work
by Pavlenko {\it et al.}~\cite{Pavlenko-2012-arXiv}, it was found that
oxygen vacancies at titanate interfaces are responsible for a
two-dimensional magnetic state and a strong orbital reconstruction.
 

In view of the previous discussion, we performed density functional
theory calculations for a number of SrTiO$_3$ slabs with various
concentrations of oxygen vacancies and analyzed the origin of the
unusual metallic state appearing at the surface of the slabs. The
paper is organized as follows.  In the next section we present the
computational details of our calculations. Section 3 is dedicated to
the discussion of the electronic structure for various slabs and
oxygen doping concentrations and finally in section 4 we summarize our
findings and relate them with the experimental observations.

\section{Computational details}

\begin{figure}
\begin{center}
\includegraphics[width=0.5\textwidth]{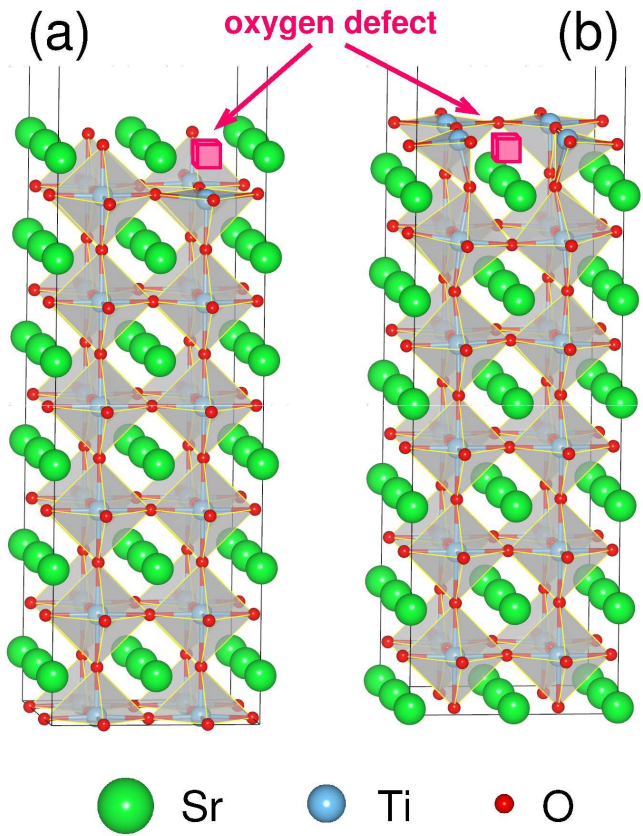}
\end{center}
\caption{Optimized structures of $2 \times 2 \times 6$ SrTiO$_3$ slabs
  with oxygen vacancy in (a) the SrO and (b) the TiO$_2$ surface
  layers.}
\label{fig:structure}
\end{figure}

We have considered stoichiometric STO slabs on a (001) surface.
Cleaving bulk STO in the [001] direction can lead either to SrO or to
TiO$_2$ surface terminations.  Guisinger {\it et al.}
\cite{Guisinger-2009-acsnano} used cross-sectional scanning tunneling
microscopy on {\it in situ} fractured SrTiO$_3$ and observed
atomically smooth terraces consisting of alternating surface
terminations of TiO$_2$ domains and SrO domains. Therefore, both kinds
of slabs with SrO and TiO$_2$ termination have been included in our
study.  The periodic slabs are separated by a vacuum layer of 20
{\AA}. In our calculations, we use the experimental lattice constant
3.905~{\AA} from the bulk SrTiO$_3$ with a cubic $P\,m\bar{3}m$
structure.~\cite{Nassau-1988-jcg} The oxygen vacancy is introduced as
a defect by removing one oxygen atom from the surface.
Fig.~\ref{fig:structure} gives an example slab illustrated by a $2
\times 2 \times 6$ supercell with (a) SrO and (b) TiO$_2$ termination.
Here and in the following, we abbreviate the more precise
``stoichiometric slab with oxygen defect in the SrO surface layer'' by
``SrO terminated slab''.

In order to clarify the electronic properties of surfaces with oxygen
defects, we have performed calculations both for perfect and for
oxygen deficient surfaces.  Supercells with sizes $1 \times 1 \times
12$, $2 \times 2 \times 4$, $2 \times 2 \times 6$, and $3 \times 3
\times 4$ were designed to represent different oxygen vacancy doping
concentrations SrTiO$_{3-x}$ with $x = 0.0833$, 0.0625, 0.0417, and
0.0278, respectively. Note that the authors of
Ref.~\onlinecite{Santander-2011-nature} reported very small oxygen
deficiency. We therefore restricted ourselves to concentrations $x <
0.1$.  The considered models are also expected to help understand the
distribution of the extra electrons introduced by the oxygen vacancy
along the in-plane ($xy$) and out-of-plane ($z$) directions.

The calculations were performed in the framework of density functional
theory (DFT).  Structure relaxations were done by considering the
Born-Oppenheimer procedure as implemented in the Vienna \textit{ab
  initio} simulation package
(VASP)~\cite{Kresse-1993-prb,Kresse-1996-cms}.  The wave functions
were represented using the projector augmented wave (PAW)
method~\cite{Blochl-1994-prb,Kresse-1999-prb} and we considered the
LDA+U functional~\cite{Liechtenstein-1995-prb} with U = 5 and J =
0.64~eV for the Ti $3d$ states~\cite{Okamoto-2006-prl} in all our
relaxations.  A plane wave cutoff of 500~eV was
used.  The Brillouin zone was sampled with a $k$-mesh of: $(4 \times 4
\times 2)$ for the $2 \times 2 \times 4$ slab, $(4 \times 4 \times 1)$
for the $2 \times 2 \times 6$ slab, $(1 \times 1 \times 1)$ for the $3
\times 3 \times 4$, and a $(8 \times 8 \times 1)$ for the $1 \times 1
\times 12$ slab. As a small surface oxygen defect concentration is not
expected to change the bulk lattice constant, we relaxed only all
internal coordinates of the slab but not the lattice parameters. Based
on the optimized structure, the electronic properties were calculated
using the all-electron full potential local orbital
(FPLO)~\cite{Koepernik-fplo1} method and we employed
both the generalized gradient approximation
(GGA)~\cite{Perdew-1996-prl} as well as the GGA+U approximation as
exchange-correlation functionals within DFT.

\section{Electronic structure}

\begin{figure}
\begin{center}
\includegraphics[width=0.4\textwidth]{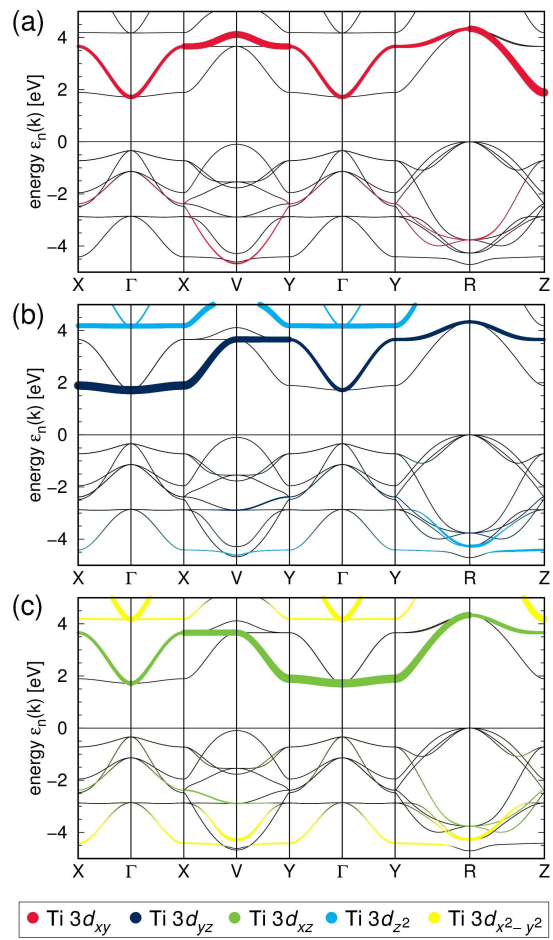}
\end{center}
\caption{Band structure with band weights of Ti for bulk STO.}
\label{fig:bw-bulk111}
\end{figure}

Bulk STO is a band insulator.  In Fig.~\ref{fig:bw-bulk111} we show
the GGA calculated band structure of bulk STO with Ti $3d$ band
weights.  The crystal field of the O octahedron surrounding Ti splits
the Ti~$3d$ bands into energetically lower $t_{2g}$ and higher $e_g$
states. We observe that the lowest unoccupied conduction band consists
of three empty Ti $t_{2g}$ bands which are degenerate at the $\Gamma$
point.  At low temperatures ($T < 105$~K) STO undergoes a structural
phase transition to a tetragonally distorted
structure~\cite{Cowley1996} and this degeneracy of the $t_{2g}$ bands
is lifted.
Fig.~\ref{fig:bw-bulk111} shows the Ti $3d$ band character analyzed in
the cartesian coordinate system, with an emphasis on the $X-\Gamma-X$
and $Y-\Gamma-Y$ paths that will become important once we break the
perfect cubic symmetry by introducing a surface.  Along the $X-\Gamma
- X$ path, the band structure consists of a weakly dispersive
heavy-mass Ti $3d_{yz}$ band and a pair of degenerate strongly
dispersive light-mass Ti $3d_{xy}$ and Ti $3d_{xz}$ bands; along
$Y-\Gamma-Y$, electrons in the Ti $3d_{xz}$ band are heavy while those
in degenerate $3d_{xy}$ and Ti $3d_{yz}$ are light.

In the following we will present a comparative analysis of various STO
slabs and discuss the changes in the electronic properties observed
when considering (i) undoped versus doped STO (ii) unrelaxed versus
relaxed STO slab structures (iii) GGA versus GGA+U calculations, (iv)
SrO-terminated versus TiO$_2$-terminated slabs and (v) different
oxygen-vacancy concentrations.

\begin{figure}
\begin{center}
\includegraphics[angle=-90,width=0.45\textwidth]{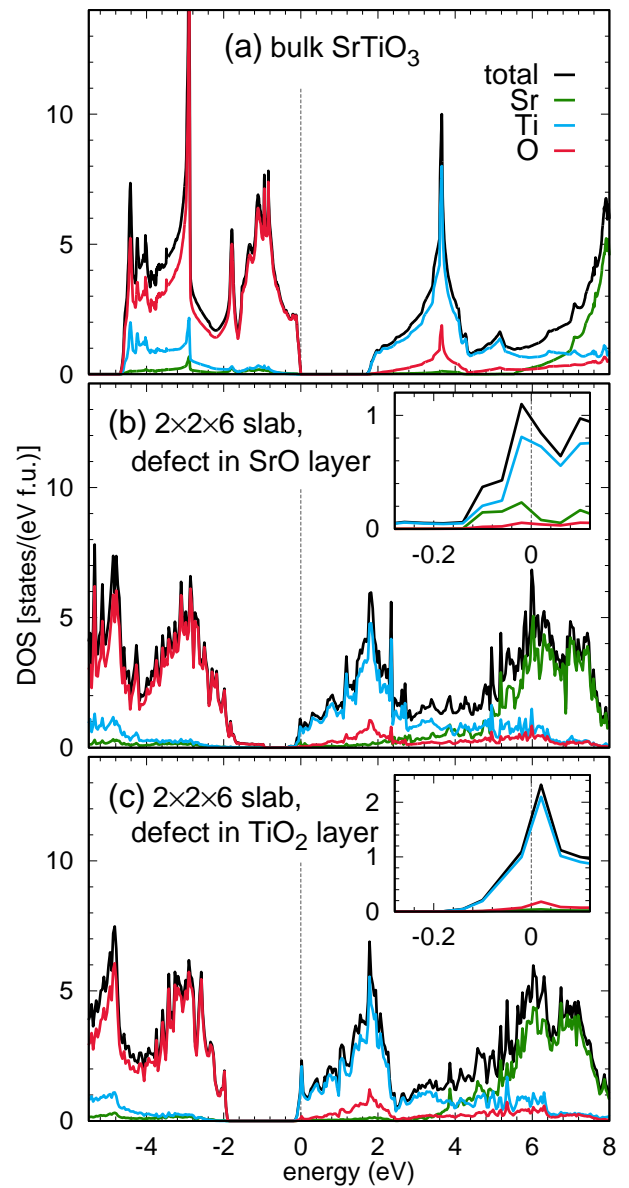}
\end{center}
\caption{Total and partial density of states for bulk STO and cleaved
  slab with oxygen vacancy.(a) bulk STO; (b) $2 \times 2 \times 6$
  with SrO termination; (c) $2 \times 2 \times 6$ with TiO$_2$
  termination. The insert figures are the zoom in of the bottom of
  conduction band.}
\label{fig:dos111-226}
\end{figure}

In Fig.~\ref{fig:dos111-226}~(a) we present the GGA calculated density
of states (DOS) for bulk STO with no oxygen vacancies compared to
relaxed $2 \times 2 \times 6$ SrO and TiO$_2$ terminated slabs
(Figs.~\ref{fig:dos111-226}~(b) and (c)) with an oxygen-vacancy
concentration of $x=0.0417$.  We observe that the extra electrons
coming from the oxygen vacancy occupy the bottom of the conduction
band with mostly Ti $3d$ character, and a metallic state is obtained
(compare Fig.~\ref{fig:dos111-226} (a) with Fig.~\ref{fig:dos111-226}
(b) and (c)).  One should mention here that the (relaxed) STO slabs
without oxygen deficiencies show, as in the case of bulk STO, a band
insulating state.~\cite{Muthu_2011}

\begin{figure}
\begin{center}
\includegraphics[width=0.4\textwidth]{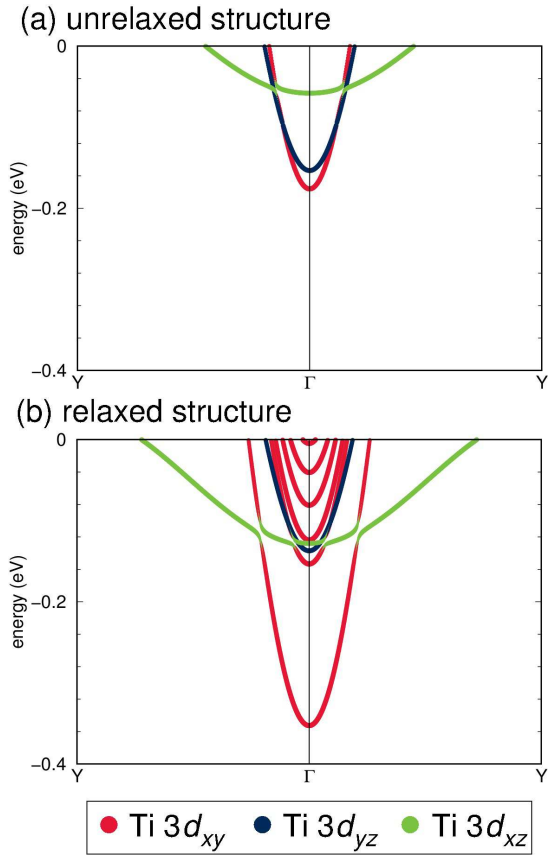}
\end{center}
\caption{Band structure obtained within GGA for a $1 \times 1 \times
  12$ slab ($x=0.0833$) with TiO$_2$ termination.  (a) unrelaxed
  structure (b) relaxed structure.}
\label{rel_unrel}
\end{figure}

Essential for a reliable description of these systems is the
structural relaxation of the slab geometry.  In Fig.~\ref{rel_unrel}
we show Ti $3d$ orbital-resolved band structures near the Fermi level
for unrelaxed and relaxed geometries of the $1 \times 1 \times 12$
slab with an oxygen-vacancy concentration of $x= 0.0833$ for a TiO$_2$
terminated surface.  We observe a significant surface reconstruction
after structure relaxation.  While relaxation has little effect on the
$3d_{xz}$ band, the splitting between $3d_{xy}$ and $3d_{xz}$ is
considerable after relaxation, and the $3d_{xz}$ band becomes slightly
less heavy ($m^*=7.9m_e$ (relaxed) against $m^*=8.4m_e$ (unrelaxed)).
This is in contrast to previous calculations that were performed
without relaxing the structure~\cite{Santander-2011-nature}.

\begin{figure} 
\begin{center}
\includegraphics[width=0.5\textwidth]{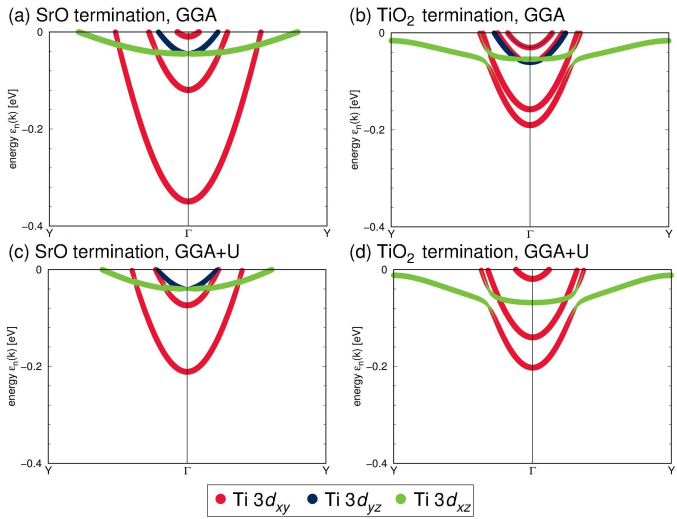}
\end{center}
\caption{Band structure obtained within GGA and GGA+U for a
  $2 \times 2 \times 4$ slab ($x=0.0625$).
 (a) GGA for SrO termination; (b) GGA for TiO$_2$ termination;  (c) GGA+U
  for SrO termination; (d) GGA+U for
  TiO$_2$ termination.}
\label{fig:bw224-GGA-GGAU}
\end{figure}

In order to analyze the importance of correlation effects (at least at
the mean-field level as implemented in GGA+U) in the 
oxygen-deficient slabs, we show in Fig.~\ref{fig:bw224-GGA-GGAU} the
band structure for SrO and TiO$_2$-terminated $2 \times 2 \times 4$
slabs with an oxygen-vacancy concentration of $x=0.0625$ near the
Fermi level region along the $\Gamma - Y$ path both within GGA and the
GGA+U approach with $U= 5$~eV and $J=0.64$~eV.  Apart from a narrowing
of the bandwidth in the GGA+U calculations with respect to the GGA
results (compare Figs.~\ref{fig:bw224-GGA-GGAU} (c) and (d) with
Figs.~\ref{fig:bw224-GGA-GGAU} (a) and (b) respectively) the main
features of the metallic state are not influenced. We will concentrate
in the following on GGA calculations in order to save some
computational effort but also in order to sidestep an unnecessary
arbitrariness arising from the choice of $U$ and $J$ values.

\begin{figure}
\begin{center}
\includegraphics[width=0.4\textwidth]{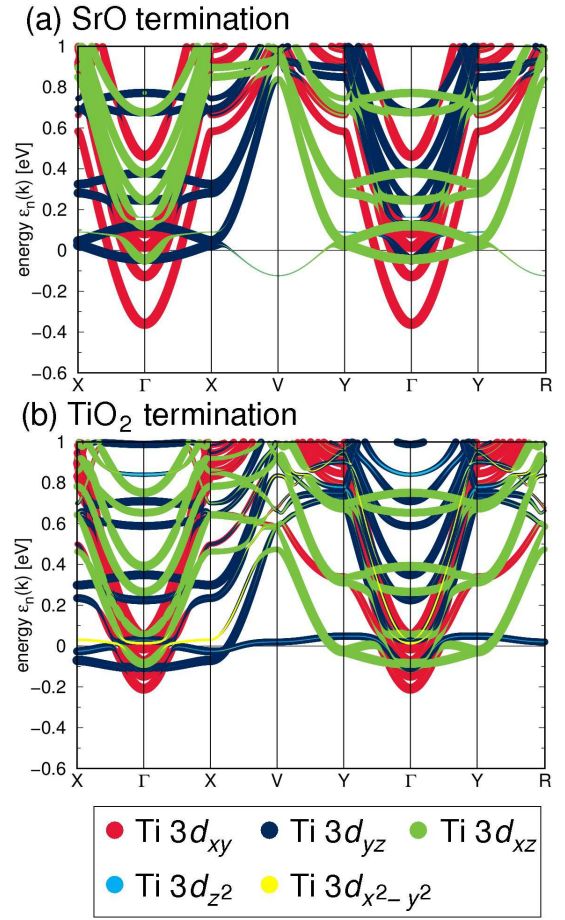}
\end{center}
\caption{Calculated band weights of Ti $3d$ for the $2 \times 2 \times
  6$ slab with oxygen vacancy. (a) SrO termination; (b) TiO$_2$
  termination.}
\label{fig:226fullbw}
\end{figure}

\begin{figure}
\begin{center}
\includegraphics[width=0.5\textwidth]{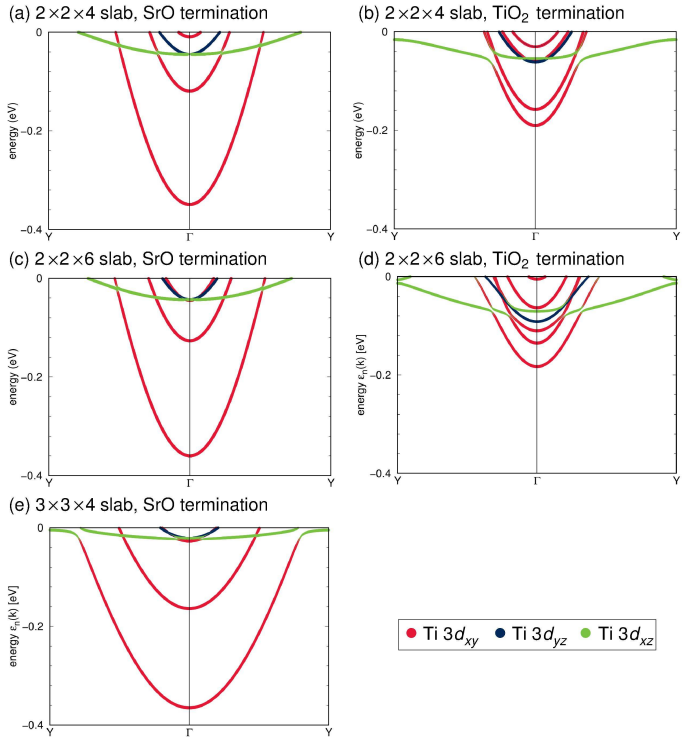}
\end{center}
\caption{Calculated band structures for different (relaxed) slabs $2
  \times 2 \times 4$, and $2 \times 2 \times 6$, and $3 \times 3
  \times 4$ with SrO termination (left panels) and TiO$_2$ termination
  (right panels).  (a) $2 \times 2 \times 4$ slab with SrO
  termination, (b) $2 \times 2 \times 4$ slab with TiO$_2$
  termination, (c) $2 \times 2 \times 6$ slab with SrO termination,
  (d) $2 \times 2 \times 6$ slab with TiO$_2$ termination, (e) $3
  \times 3 \times 4$ with SrO termination.
}
\label{fig:bw-1112-224-226-334}
\end{figure}

Next, we analyze the differences in electronic behavior of
oxygen-deficient SrO-terminated slabs and TiO$_2$-terminated slabs.
In Fig~\ref{fig:226fullbw}~(a) and (b) we present the band structure
for a relaxed $2\times 2 \times 6$ slab with an oxygen-vacancy
concentration of $x=0.0417$ for the two terminations.  Since the slab
breaks some symmetries present in the bulk, the electronic structure
splits into a ladder of subbands caused by the Ti atoms becoming
non-equivalent.  For both terminations, the Ti bands that were
unoccupied conduction bands in the bulk (see
Fig.~\ref{fig:bw-bulk111}), are now partially filled with the extra
electrons gained from the oxygen vacancies, as mentioned above.  The
lowest sub-bands come from in-plane Ti $3d_{xy}$ states largely
localized on the surface TiO$_2$ layer (TiO$_2$ termination) or on the
layer right below the surface layer (SrO termination). The slab
termination affects to some extent the band splitting in the $3d_{xy}$
bands at $\Gamma$.  For the same oxygen-vacancy doping, TiO$_2$
terminated slabs lead to the occupation of more Ti $3d_{xy}$ bands
than the SrO terminated slabs; this is simply due to the number of
TiO$_2$ layers significantly affected by the defect which is three in
case of TiO$_2$ termination versus two in case of SrO termination.  In
fact, a closer look at the slab structures reveals that the
oxygen-vacancy environment in TiO$_2$-terminated slabs is more
distorted than in SrO-terminated slabs.  This trend can also be
observed at different oxygen-vacancy concentrations in
Figs.~\ref{fig:bw-1112-224-226-334} (a)-(e) where the band structure
is shown for relaxed $2 \times 2 \times 4$, $2 \times 2 \times 6$ and
$3 \times 3 \times 4$ slabs ($x=0.0625$, $x= 0.0417$ and $x=0.0278$
respectively) for both SrO and TiO$_2$ terminated surfaces. Note that
the $3 \times 3 \times 4$ slabs are computationally very demanding so
that we restricted the analysis to the SrO termination.  Also, while
it appears that the Ti $3d_{xy}$ parabolas are wider for the $3 \times
3 \times 4$ slab (Fig.~\ref{fig:bw-1112-224-226-334} (e)) compared to
the $2 \times 2 \times 4$ and $2 \times 2 \times 6$ slabs
(Fig.~\ref{fig:bw-1112-224-226-334} (a)-(d)), this is only an effect
of the different Brillouin zones; while the $Y=(0,\nicefrac{1}{2},0)$
point is at a distance of $0.40$~{\AA}$^{-1}$ from the $\Gamma$ point
in the $2 \times 2$ supercells, it is at a distance of
$0.27$~{\AA}$^{-1}$ for the $3 \times 3$ supercell.

Comparing these band structures to ARPES
observations~\cite{Santander-2011-nature,Meevasana-2011-nmat}, we find
good agreement. While Meevasana {\it et al.} see only the light Ti
$3d_{xy}$ bands~\cite{Meevasana-2011-nmat}, Santander-Syro {\it et
  al.} detect also the shallow heavy bands deriving from Ti
$3d_{xz}$~\cite{Santander-2011-nature}.

We have also investigated the role of the higher lying $e_g$
states~\cite{Pavlenko-2012-arXiv} after introducing oxygen vacancies
and observe that in contrast to the LAO/STO system, when an oxygen
defect is introduced on the bare STO surface, the $e_g$ states of the
Ti atoms neighboring the oxygen vacancy are hardly affected.

\subsection{ Two-dimensional electron gas}

\begin{figure}
\begin{center}
\includegraphics[width=0.4\textwidth]{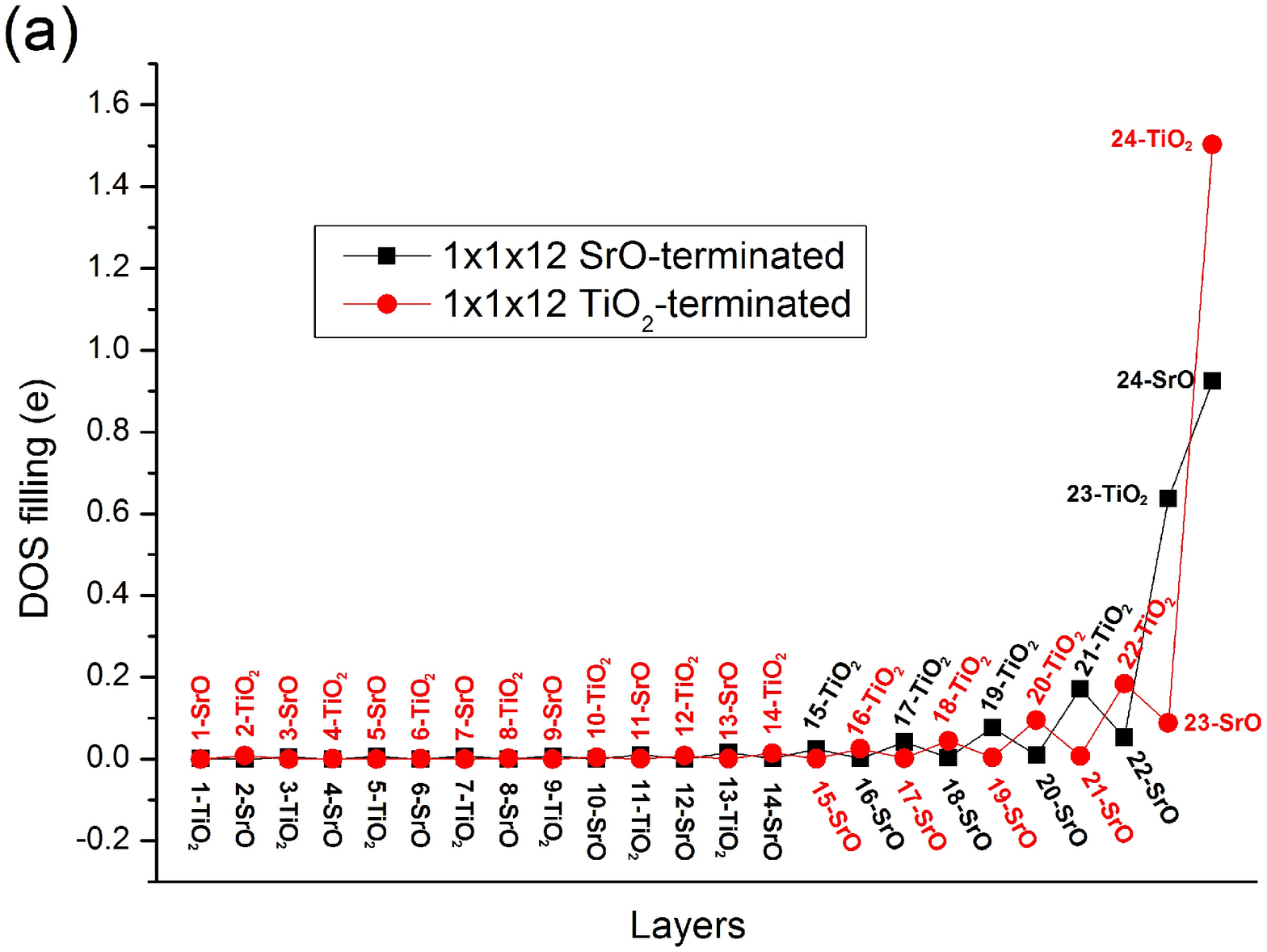}
\includegraphics[width=0.4\textwidth]{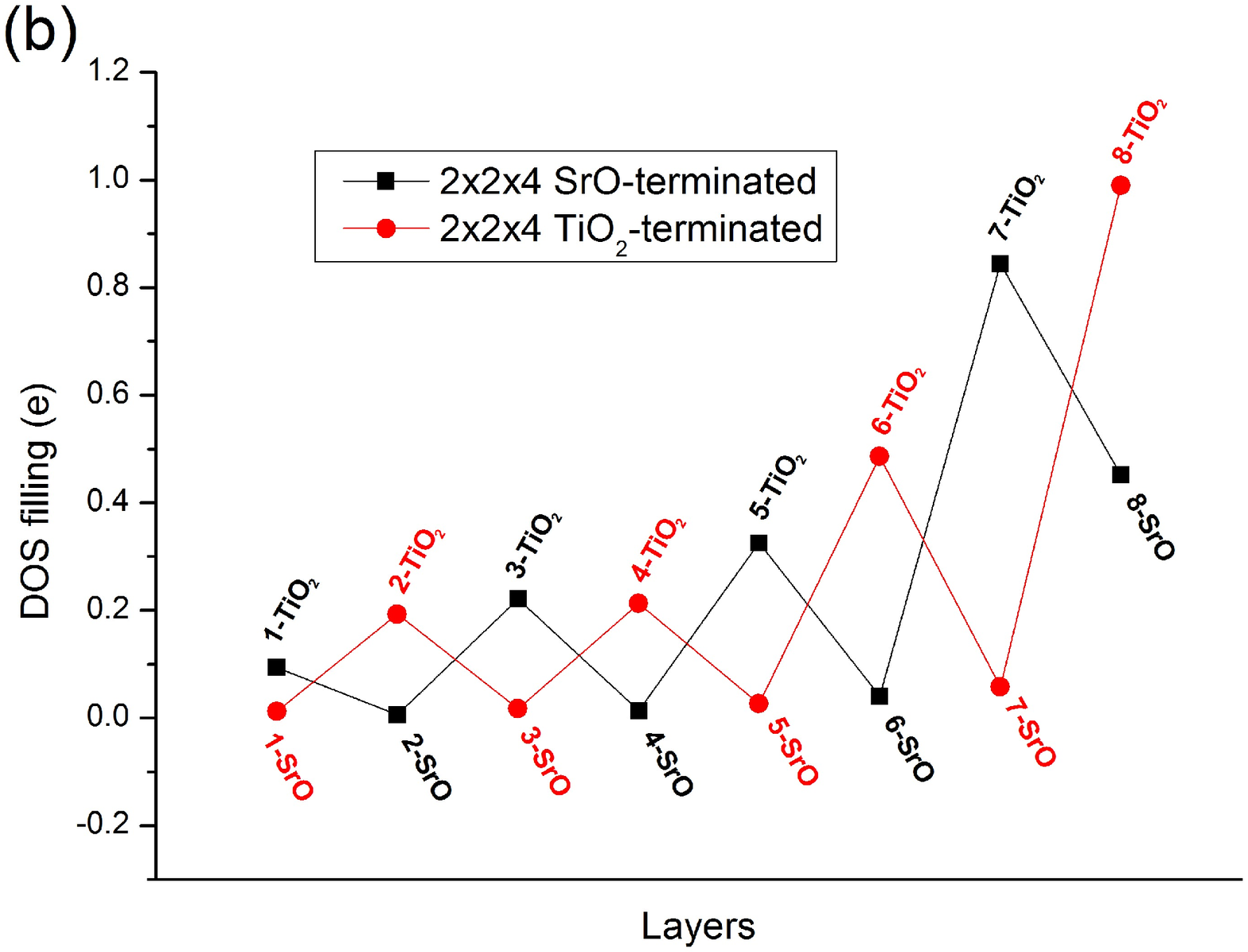}
\includegraphics[width=0.4\textwidth]{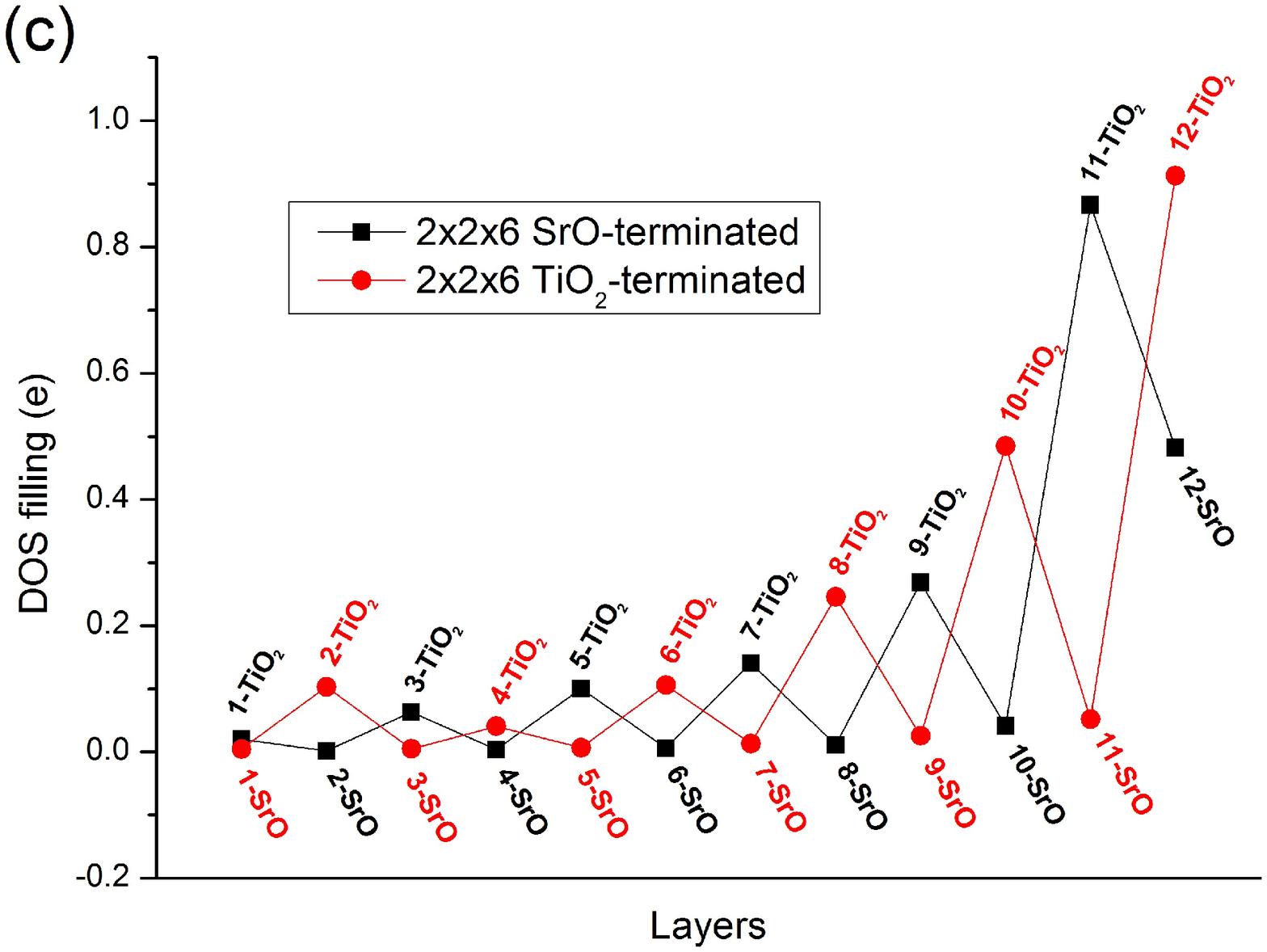}
\end{center}
\caption{DOS filling of gap states for layer decomposed PDOS of SrO
  and TiO$_2$ terminated slabs.}
\label{fig:dosfilling}
\end{figure}

Aiura {\it et al.}\cite{Aiura-2002-surfs} studied the behavior of the
metallic state of lightly electron-doped bulk SrTiO$_3$ with
angle-integrated ultraviolet photoemission spectroscopy (UPS) and
angle-resolved photoemission spectroscopy.  They observed a
metallic state with a sharp Fermi cut-off in the bulk band gap by
doping electron carriers.  By slight oxygen exposure, the UPS spectra
showed a rapid decrease in the spectral intensity of the metallic
state followed by a sudden energy shift of the valence band.
Therefore, these spectral features in the bulk system are believed to
be related to the oxygen deficiency.  The difference between the
metallic state for electron-doped bulk STO and electron-doped STO
surfaces is the localization of the itinerant electrons on the surface
in the latter case, as we find in our calculations.

In the following we analyze the confinement of the electronic charge
in the oxygen-deficient slabs.  Figure~\ref{fig:dosfilling} displays
the layer-decomposed electron filling of the gap states close to the
Fermi level for $1 \times 1 \times 12$ ($x=0.0833$), $2 \times 2
\times 4$ ($x=0.0625$) and $2 \times 2 \times 6$ ($x=0.0417$) slabs
with both SrO and TiO$_2$ termination as a funtion of depth.  The
layers are numbered from the bottom of the slab and the oxygen vacancy
has been introduced on the top layer.
 
Analyzing Fig.~\ref{fig:dosfilling} (a), we observe that the number of
electrons doped into the system due to the oxygen defect decreases
from the surface to the bottom of the slab, and there is a dramatic
drop after two layers for SrO termination and after only one layer for
TiO$_2$ termination.  This indicates that indeed a two-dimensional
electron gas (2DEG) forms at the surface.  The charge resulting from
surface-localized oxygen vacancies forms a narrow distribution that is
peaked at the surface and represents a narrow 2DEG in the 3D
crystal. The charge distribution of Fig.~\ref{fig:dosfilling} (a)
agrees well with the approximate solution of the Poisson equation
which was used in Ref.~\onlinecite{Meevasana-2011-nmat} to
characterize the observed 2DEG.  The number of electrons effectively
drops to zero after ten layers (if we count TiO$_2$ and SrO layers
separately), which means that the system recovers its bulk properties
within 5 unit cells below the surface. With the exception of the top
layer in the SrO terminated slabs, excess electrons accumulate mostly
on TiO$_2$ layers. Nevertheless, the $1 \times 1 \times 12$ slab has
an excessively large oxygen-defect density at the surface with respect
to the reported experimental oxygen
deficiency~\cite{Santander-2011-nature}.  Even though the slab is
thick enough to see the depth effect, the generated top layer has one
oxygen vacany per one Sr atom (SrO surface termination) and per one
TiO strip (TiO$_2$ surface termination). This slab was considered here
for comparison to the calculations performed in
Ref.~\onlinecite{Santander-2011-nature}.

In Fig.~\ref{fig:dosfilling} (b), we show the corresponding
layer-decomposed electron filling of the gap states for a $2 \times 2
\times 4$ slab.  A similar size slab has also been employed as a
typical model in several previous
works.~\cite{Cai-2006-jcp,Lin-2009-prb,Luo-2004-prb,Heifets-2001-prb}
We can see that the main features are similar to the $1 \times 1
\times 12$ slab case, but the bottom layer still shows some electron
doping.  This indicates that this slab is too thin to study the
carriers in the system if the criterion that a part of the slab should
show bulk STO carrier densities is considered.
Fig.~\ref{fig:dosfilling} (c) shows a similar plot for a $2 \times 2
\times 6$ slab but in this case bulk properties are recovered at the
bottom of the slab.

For all considered slabs, the number of doping electrons in the gap
states are close to 2.0 independent of the size of the slab and type
of termination.  Except for the extreme case of the $1 \times 1 \times
12$ slab, the second TiO$_2$ layer in the SrO surface termination not
only contributes maximally to the electron distribution but also
accumulates a similar number of electrons as the 1st TiO$_2$ layer on
the TiO$_2$ surface termination slabs, while the 2nd SrO layer in the
TiO$_2$ surface termination slabs has almost no doping electrons.
This confirms that titanium oxide layers play a crucial role for the
conductivity in these systems. 

\begin{figure}
\begin{center}
\includegraphics[width=0.4\textwidth]{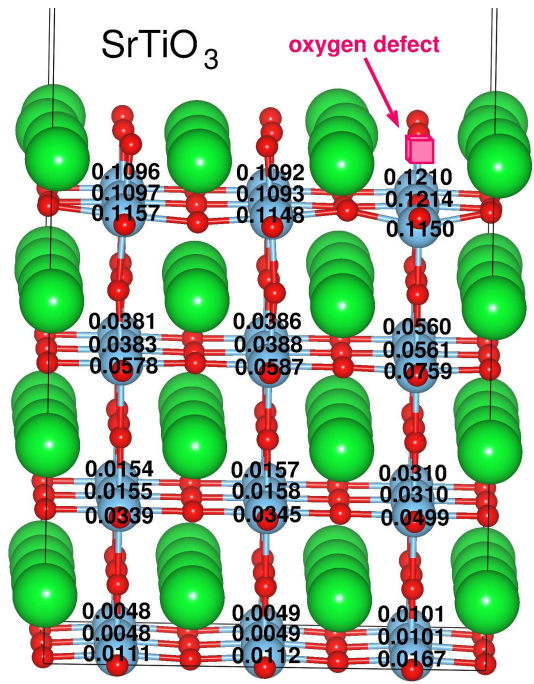}
\end{center}
\caption{Ti $3d$ occupations for the SrO terminated  $3 \times 3 \times 4$ slab.
}
\label{fig:tioccupations}
\end{figure}

In Fig.~\ref{fig:tioccupations} we show the Ti $3d$ charge occupation
for the SrO-terminated $3\times3\times4$ slab ($x = 0.0278$). We
observe that the Ti atoms directly below and diagonally adjacent to
the O defect position carry a maximal excess charge of 0.12
electrons. The other Ti sites in this layer carry a rather similar
excess charge of 0.11 electrons, making the electron doping in this
layer almost homogeneous. The next TiO$_2$ layer below shows excess
charges between 0.04 and 0.08 electrons, the third TiO$_2$ between
0.02 and 0.05 electrons. This charge distribution in the Ti $3d$
shells indicates that a description of possible correlation effects --
which is beyond the scope of the present work -- would require the
consideration of site dependent selfenergies.

\begin{figure}
\begin{center}
\includegraphics[angle=-90,width=0.45\textwidth]{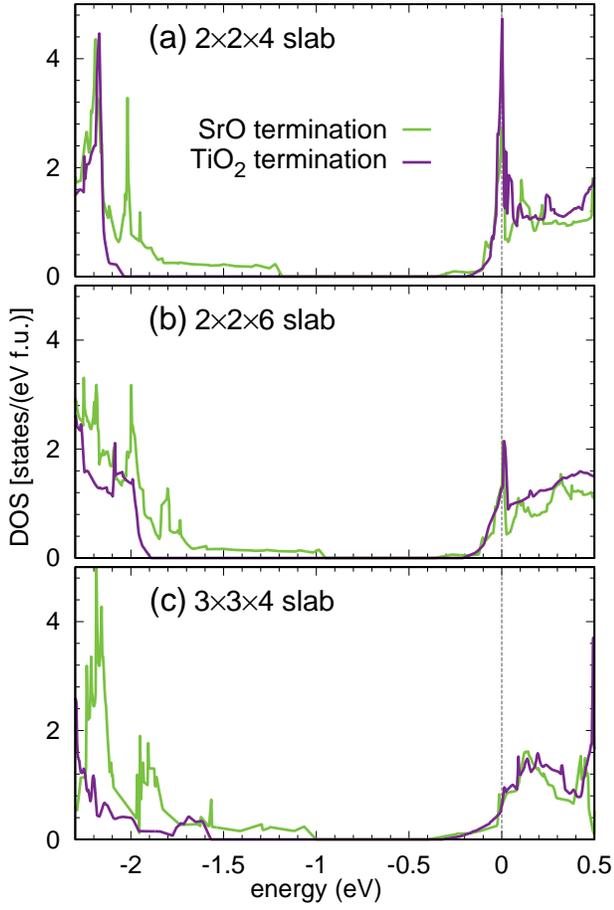}
\end{center}
\caption{ Total DOS for slab $2 \times 2 \times 4$, $2 \times 2 \times
  6$, and $3 \times 3 \times 4$, SrO and TiO$_2$ terminated normalized
  to STO formula units.
}
\label{fig:alldosscaled}
\end{figure}

A strong instability towards ferromagnetism has been recently
discussed~\cite{Pavlenko-2012-arXiv} in the context of
oxygen-deficient titanate interfaces. The density of states (DOS) for
our $2 \times 2 \times 4$, $2 \times 2 \times 6$, and $3 \times 3
\times 4$ slabs for both surface terminations
(Fig.~\ref{fig:alldosscaled}) shows a sharp peak at $\epsilon_F$
caused by the surface TiO$_2$ layer (TiO$_2$ termination) or the
TiO$_2$ layer below the SrO layer (SrO termination).  This indicates
-- according to the Stoner criterion -- a strong instability towards
ferromagnetism also in the case of a surface 2DEG.  Our data shows
that by decreasing the doping level, {\it i.e.}  increasing the slab
size, the leading edge of the valence band shifts to higher energy,
and the intensity of the sharp peak near the Fermi level decreases
respectively.

\begin{figure}
\begin{center}
\includegraphics[width=0.5\textwidth]{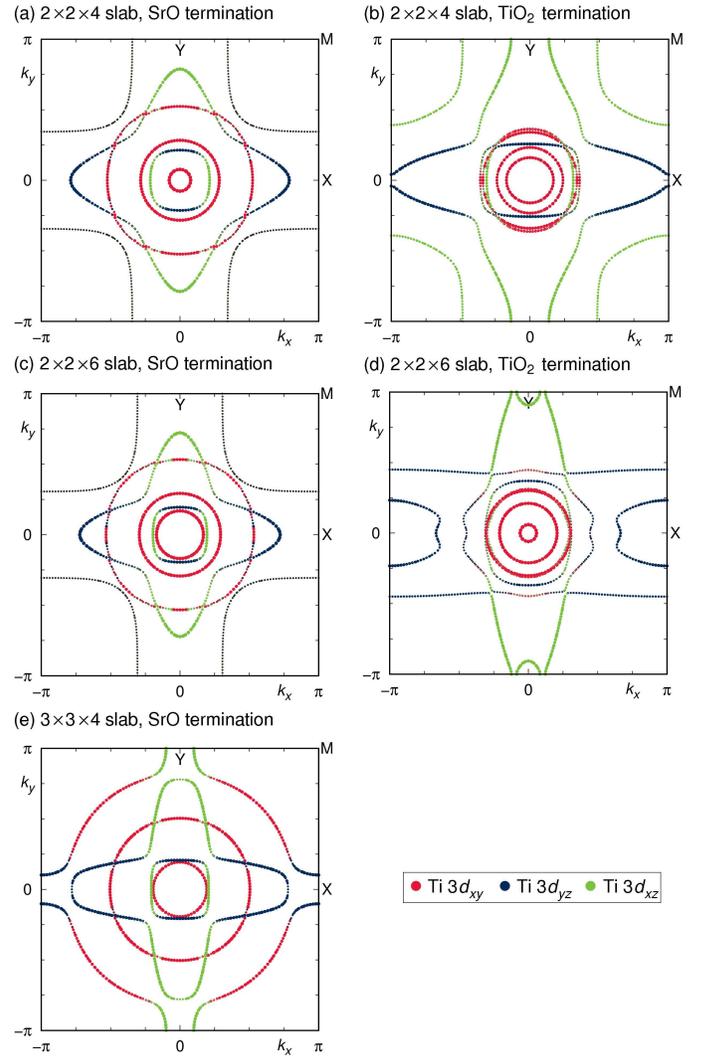}
\end{center}
\caption{Calculated Fermi surface for (a) SrO terminated $2 \times 2
  \times 4$ slab; (b) TiO$_2$ terminated $2 \times 2 \times 4$ slab;
  (c) SrO terminated $2 \times 2 \times 6$ slab; (d) TiO$_2$
  terminated $2 \times 2 \times 6$ slab, and (e) SrO terminated $3
  \times 3 \times 4$ slab.
}
\label{fig:fs-224-226-334}
\end{figure}

In Fig.~\ref{fig:fs-224-226-334} we present the calculated Fermi
surface for the considered slabs with both SrO and TiO$_2$
terminations.  The figures show that the Ti $3d_{yz}$ states have
strongly elliptical Fermi surfaces and high effective mass along $x$
compared to low effective mass along $y$, and analogously for Ti
$3d_{xz}$ with the properties along $x$ and $y$ exchanged.  Meanwhile,
Ti $3d_{xy}$ states have a circular Fermi surface and light effective
mass along $x$ and $y$. Note that in the calculations, an oxygen
defect in the SrO surface layer preserves the rotational symmetry
while a defect in the TiO$_2$ surface layer breaks it; this is why
Figs.~\ref{fig:fs-224-226-334}~(b) and (d) are not the same after
rotating by 90$^\circ$ and exchanging $3d_{yz}$ by $3d_{xz}$. In
reality, surface oxygen defects would be expected to be randomly
distributed rather than forming a regular lattice as in the
calculation.  The circular Ti $3d_{xy}$ derived Fermi surfaces compare
well to the ARPES measurements of Meevasana {\it et
  al.}~\cite{Meevasana-2011-nmat} and Santander-Syro {\it et
  al.}~\cite{Santander-2011-nature}. Santander-Syro {\it et al.}  also
extract signatures of the elliptical Ti $3d_{yz}/d_{xz}$ Fermi surface
pockets~\cite{Santander-2011-nature} as found in our calculations. The
elliptical Fermi surface pockets have also been observed by Chang {\it
  et al.}~\cite{Chang2010} in TiO$_2$ terminated STO single crystals
where oxygen vacancies were introduced by annealing in ultra high
vaccuum.

Finally, we present in Table~\ref{tab:effectivemass} the effective
orbital-resolved electronic masses $m^*$ for the various slabs
calculated from
\begin{equation}
m^* = \bigg(\frac{1}{\hbar^2}\frac{d^2 \epsilon_k}{dk^2}\bigg)^{-1},
\end{equation}
where $\hbar$ is the Planck constant and $\epsilon_k$ is the
$k$-resolved band energy. We find very good agreement with the
experimental observations for the light effective masses ($m^\ast =
0.5 \sim 0.7 m_e$)~\cite{Santander-2011-nature,Meevasana-2011-nmat}.
The heavy effective masses are more difficult to determine
experimentally and are in the range $m_y^*\approx
10m_e-20m_e$~\cite{Santander-2011-nature}; our values of $m^*\in
[7.5m_e,9.3m_e]$ are in reasonable agreement with
Ref.~\onlinecite{Santander-2011-nature} and in very good agreement
with the value $m^*/m_e=7.0$ measured in Ref.~\onlinecite{Chang2010}.

\begin{table}
  \caption{ Calculated orbital-resolved effective masses $m^*$ for
    different oxygen-deficient slab sizes and surface terminations.
    The error bar for Ti $3d_{xy}$ is $\pm$0.01, and for Ti $3d_{xz}$
    is $\pm$1. }
\begin{tabular}{ccccccc}
\hline\hline
&&$1 \times 1 \times 12$&$2 \times 2 \times 4$&$2 \times 2 \times 6$&$3 \times 3 \times 4$\\\hline
$x$ (STO$_{3-x}$)&&0.0833&0.0625&0.0417&0.0278\\\hline
$m^*/m_e$ ($d_{xy}$, SrO) &&0.46&0.43&0.43&0.43\\
$m^*/m_e$ ($d_{xy}$, TiO$_2$) &&0.42&0.40&0.47&\\\hline
$m^*/m_e$ ($d_{xz}$, SrO) &&8.6&8.3&7.5&9.0\\
$m^*/m_e$ ($d_{xz}$, TiO$_2$) &&7.9&9.3&8.7&\\
\hline\hline
\end{tabular}
\label{tab:effectivemass}
\end{table}

\begin{figure}
\includegraphics[width=0.5\textwidth]{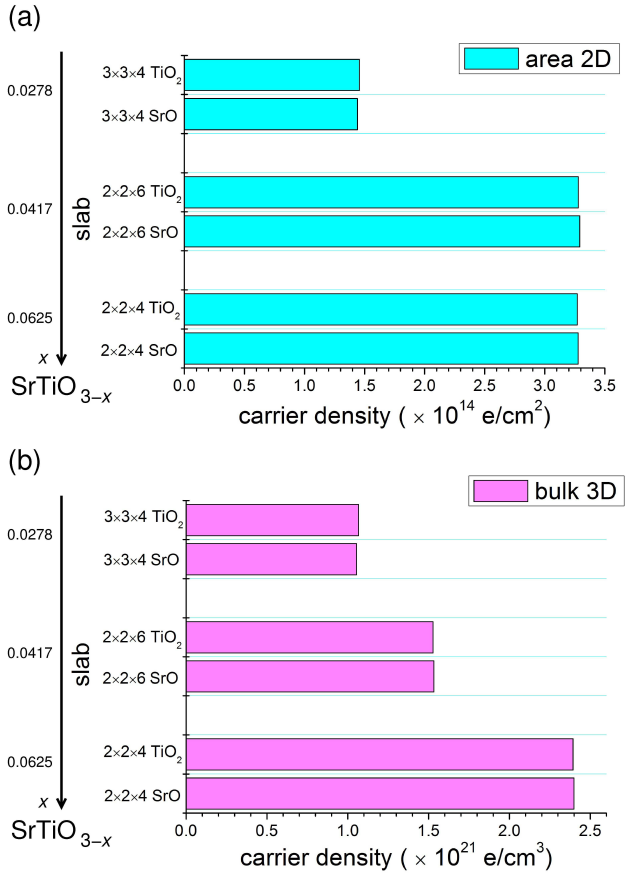}
\caption{Calculated carrier density for the SrO and TiO$_2$ terminated
  $2 \times 2 \times 6$ slabs; (a) carrier density per surface area;
  (b) carrier density per volume.}
\label{fig:carrier-dens}
\end{figure}

In Fig.~\ref{fig:carrier-dens} we compare the carrier density of our
slabs in terms of number of carriers per unit cell area versus number
of carriers per unit cell volume.  Our two-dimensional carrier density
compares well with the experimental data ($n_{2D} = 2
\times10^{14}$~cm$^{-2}$)~\cite{Santander-2011-nature,Meevasana-2011-nmat}. We
also observe that the number of carriers per area doesn't scale with
the oxygen vacancy concentration as it happens for the bulk electron
doping $n_{3D}$.  This result strongly indicates that the observed
metallic state doesn't correspond to bulk electron doping and the 2DEG
on the surface is stable independent of oxygen-vacancy concentration
and bulk doping density.

\section{Summary}

By employing density functional theory we have studied the electronic
properties of various oxygen-deficient STO surface slabs SrTiO$_{3-x}$
for $x < 0.1$ with special emphasis on the role of (i) slab structure
relaxation, (ii) electron correlation at the mean-field level, (iii)
surface termination and (iv) oxygen-vacancy concentration. We detect
a significant surface reconstruction after including oxygen vacancies
and we observe the formation of a metallic state only after
introduction of oxygen vacancies.  The charge carriers --
independently of the oxygen concentration -- are strongly localized at
the surface and deplete rapidly within a few layers from the surface,
which is an indication of the formation of a two-dimensional electron
gas.  Our calculated Fermi surfaces, effective masses and
two-dimensional carrier densities show very good agreement with
experiment.

The presented calculations show that the presence of oxygen vacancies
is essential for the formation of a 2DEG at the surface of STO and set
a basis for further detailed investigations on oxygen-deficient
surfaces and interfaces beyond density functional theory.

\section{Acknowledgements}
The authors would like to thank Ralph Claessen, Michael Sing and Thilo
Kopp for useful discussions and gratefully acknowledge financial
support by the Deutsche Forschungsgemeinschaft (DFG) through grant FOR
1346, the Beilstein-Institut, Frankfurt/Main, Germany, within the
research collaboration NanoBiC as well as by the Alliance Program of
the Helmholtz Association (HA216/EMMI).  The generous allotment of
computer time by CSC-Frankfurt and LOEWE-CSC is gratefully
acknowledged.
 
\appendix

\section{Additional electronic structure analysis}

\begin{figure}
\includegraphics[width=0.5\textwidth]{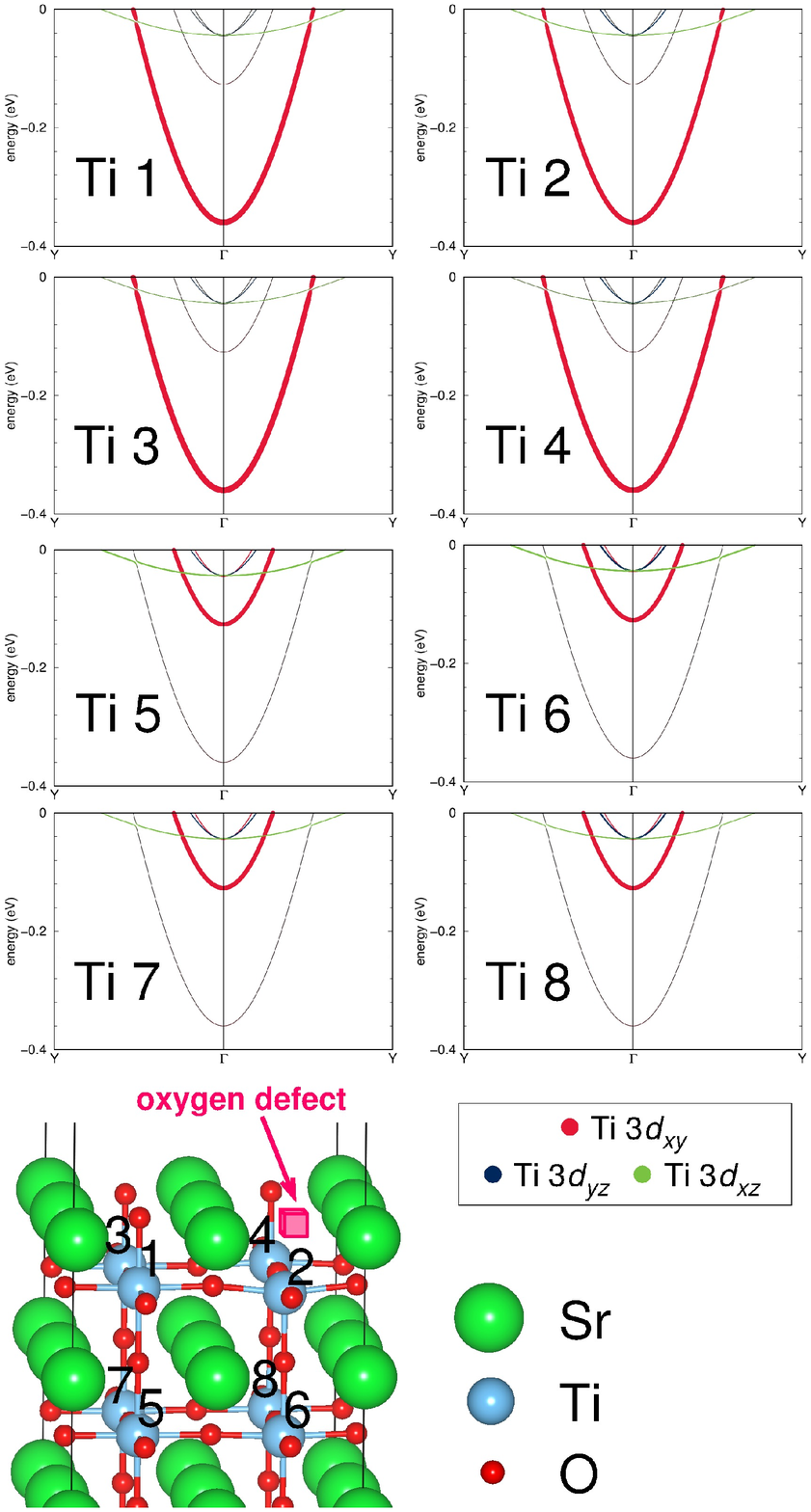}
\caption{Calculated band weights of Ti $3d$ for the $2 \times 2 \times
  6$ slab with oxygen vacancy and SrO
  termination. The contribution of each Ti in the two topmost TiO$_2$ layers is shown separately.
}
\label{fig:226banddetailSrO}
\end{figure}

\begin{figure}
\includegraphics[width=0.5\textwidth]{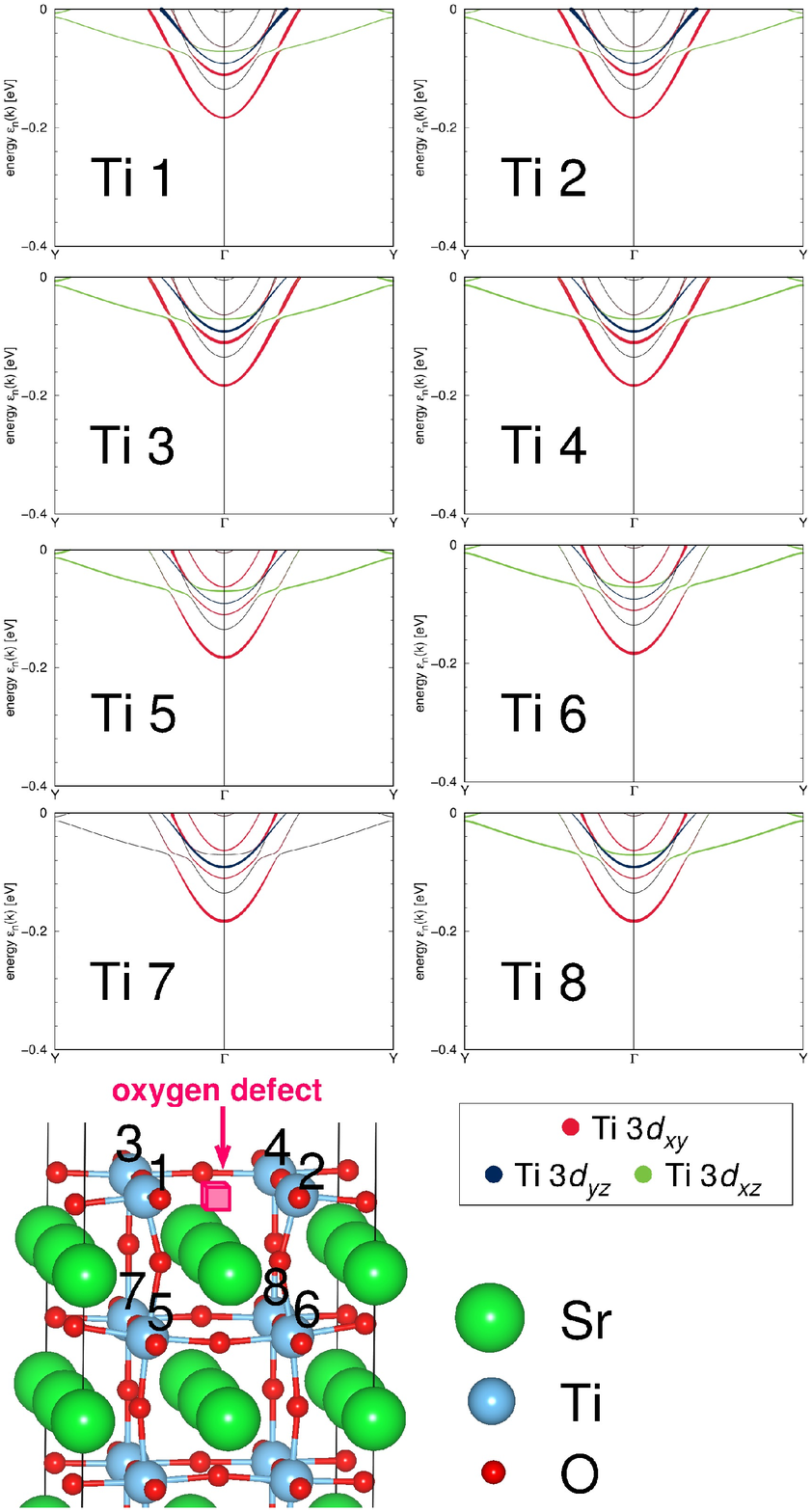}
\caption{Calculated band weights of Ti $3d$ for the $2 \times 2 \times
  6$ slab with oxygen vacancy and TiO$_2$
  termination. The contribution of each Ti in the two topmost TiO$_2$ layers is shown separately.
}
\label{fig:226banddetailTiO2}
\end{figure}

We have investigated the contribution to the band structure of the Ti
atoms in the surface layers containing the oxygen vacancy.  In
Figures~\ref{fig:226banddetailSrO} and \ref{fig:226banddetailTiO2}, we
show this analysis for $2 \times 2 \times 6$ slabs with SrO and
TiO$_2$ termination, respectively. For the SrO terminated slab, the
four Ti atoms closest to the surface (Ti$\,$1-Ti$\,$4) almost equally
contribute to the lowest Ti $3d_{xy}$ band. In the second TiO$_2$
layer, contributions from Ti$\,$5-Ti$\,$8 are again very similar, but
this time they are responsible for the second lowest Ti $3d_{xy}$ band
and to some extent contribute to the Ti $3d_{xz}$ and $3d_{yz}$ bands.
For the TiO$_2$ terminated slab (see
Fig.~\ref{fig:226banddetailTiO2}), the fact that the defect breaks the
rotational symmetry in the surface TiO$_2$ layer is clearly visible:
Ti$\,$1 and Ti$\,$2, which are adjacent to the oxygen defect, show
similar contributions to the band structure; these contributions are
different from Ti$\,$3 and Ti$\,$4, which didn't lose a bond due to
the formation of the defect. Overall, in the TiO$_2$ termination case,
the occupied bands are less easily assigned to subsequent TiO$_2$
layers than in the case of SrO termination.

\begin{thebibliography}{99}

\bibitem{Santander-2011-nature}
A. F. Santander-Syro, O. Copie, T. Kondo, F. Fortuna, S. Pailh\`{e}s, R. Weht, X. G. Qiu, F. Bertran, A. Nicolaou, A. Taleb-Ibrahimi, P. Le F\`{e}vre, G. Herranz, M. Bibes, N. Reyren, Y. Apertet, P. Lecoeur, A. Barth\'{e}l\'{e}my, and M. J. Rozenberg, Nature (London) {\bf 469}, 189 (2011).

\bibitem{Ohtomo-2004-nature}
A. Ohtomo and H. Y. Hwang, Nature (London) {\bf 427}, 423 (2004). 

\bibitem{Eckstein-2007-nmat}
J. N. Eckstein, Nat. Mater. {\bf 6}, 473 (2007). 

\bibitem{Reyren-2007-science}
N. Reyren, S. Thiel, A. D. Caviglia, L. Fitting Kourkoutis, G. Hammerl, C. Richter, C. W. Schneider, T. Kopp, A.-S. R{\" u}etschi, D. Jaccard, M. Gabay, D. A. Muller, J.-M. Triscone, and J. Mannhart, Science {\bf 317}, 1196 (2007). 

\bibitem{Popovic-2008-prl}
Z. S. Popovi{\' c}, S. Satpathy, and R. M. Martin, Phys. Rev. Lett. {\bf 101}, 256801 (2008).

\bibitem{Basletic-2008-nmat}
M. Basletic, J.-L. Maurice, C. Carr\'{e}t\'{e}ro, G. Herranz1, O. Copie, M. Bibes, \'{E}. Jacquet, K. Bouzehouane, S. Fusil,  A. Barth\'{e}l\'{e}my, Nat. Mater. {\bf 7}, 621 (2008). 

\bibitem{Cen-2009-science}
C. Cen, S. Thiel, J. Mannhart, and J. Levy, Science {\bf 323}, 1026 (2009). 

\bibitem{Janicka-2009-prl}
K. Janicka, J. P. Velev, and E. Y. Tsymbal, Phys. Rev. Lett. {\bf 102}, 106803 (2009). 

\bibitem{Li-2011-science}
L. Li, C. Richter, S. Paetel, T. Kopp, J. Mannhart, and R. C. Ashoori, Science {\bf 332}, 825 (2011).

\bibitem{Arras-2012-prb}
R. Arras, V. G. Ruiz, W. E. Pickett, and R. Pentcheva, Phys. Rev. B {\bf 85}, 125404 (2012).

\bibitem{Zhong_2012} Z. Zhong, P. Wissgott, K. Held, and
  G. Sangiovanni, Europhys. Lett. {\bf 99}, 37011 (2012).

\bibitem{Huijben-2006-nmat}
M. Huijben, G. Rijnders, D. H. A. Blank, S. Bals, S. van Aert, J. Verbeeck, G. van Tendeloo, A. Brinkman, and H. Hilgenkamp, Nat. Mater. {\bf 5}, 556 (2006). 

\bibitem{Nakagawa-2005-nmat}
N. Nakagawa, H. Y. Hwang, and D. A. Muller, Nat. Mater. {\bf 5}, 204 (2006). 

\bibitem{Thiel-2006-science}
S. Thiel, G. Hammerl, A. Schmehl, C. W. Schneider, and J. Mannhart, Science {\bf 313}, 1942 (2006).

\bibitem{Meevasana-2011-nmat}
W. Meevasana, P. D. C. King, R. H. He, S-K. Mo, M. Hashimoto, A. Tamai, P. Songsiriritthigul, F. Baumberger, and Z-X. Shen, Nat. Mater. {\bf 10}, 114 (2011).

\bibitem{Pavlenko-2012-arXiv} N. Pavlenko, T. Kopp, E. Y. Tsymbal,
  J. Mannhart, G. A. Sawatzky, arXiv:1204.4711v1.

\bibitem{Guisinger-2009-acsnano} N. P. Guisinger, T. S. Santos,
  J. R. Guest, T-Y. Chien, A. Bhattacharya, J. W. Freeland, and
  M. Bode, ACS Nano {\bf 3}, 4132 (2009).

\bibitem{Nassau-1988-jcg}
K. Nassau and A. E. Miller, J. Cryst. Growth {\bf 91}, 373 (1988).

\bibitem{Kresse-1993-prb}
G. Kresse and J. Hafner, Phys. Rev. B {\bf 47}, 558 (1993).

\bibitem{Kresse-1996-cms}
G. Kresse and J. Furthm\"uller, Comput. Mater. Sci. {\bf 6}, 15 (1996).

\bibitem{Blochl-1994-prb}
P. E. Bl\"ochl, Phys. Rev. B {\bf 50}, 17953 (1994).

\bibitem{Kresse-1999-prb}
G. Kresse and D. Joubert, Phys. Rev. B {\bf 59}, 1758 (1999).

\bibitem{Liechtenstein-1995-prb}
A. I. Liechtenstein, V. I. Anisimov, and J. Zaanen, Phys. Rev. B {\bf 52}, R5467 (1995).

\bibitem{Okamoto-2006-prl}
S. Okamoto, A. J. Millis, and N. A. Spaldin, Phys. Rev. Lett. {\bf 97}, 056802 (2006).


\bibitem{Koepernik-fplo1}
K. Koepernik and H. Eschrig, Phys. Rev. B {\bf 59}, 1743 (1999),
http://www.FPLO.de.

\bibitem{Perdew-1996-prl}
J. P. Perdew, K. Burke, and M. Ernzerhof, Phys. Rev. Lett. {\bf 77}, 3865 (1996).

\bibitem{Cowley1996}
R. A. Cowley, Phil. Trans. R. Soc. A {\bf 354}, 2799 (1996).

\bibitem{Muthu_2011} see for instance 
K. Muthukumar, I. Opahle, J. Shen, H.O. Jeschke, and R. Valent\'\i,
Phys. Rev. B {\bf 84}, 205442 (2011) for a detailed comparison
of the electronic properties of bulk versus slab geometries in
the context of SiO$_2$.

\bibitem{Aiura-2002-surfs}
Y. Aiura, I. Hase, H. Bando, T. Yasue, T. Saitoh, and D. S. Dessau, Surf. Sci. {\bf 515}, 61 (2002).

\bibitem{Cai-2006-jcp}
M. Cai, Y. Zhang, G. Yang, Z. Yin, M. Zhang, W. Hu, and Y. Wang, J. Chem. Phys. {\bf 124}, 174701 (2006).

\bibitem{Lin-2009-prb}
F. Lin, S. Wang, F. Zheng, G. Zhou, J. Wu, B.-L. Gu, and W. Duan, Phys. Rev. B {\bf 79}, 035311 (2009). 

\bibitem{Heifets-2001-prb}
E. Heifets, R. I. Eglitis, E. A. Kotomin, J. Maier, and G. Borstel, Phys. Rev. B {\bf 64}, 235417 (2001).

\bibitem{Luo-2004-prb}
W. Luo, W. Duan, S. G. Louie, and M. L. Cohen, Phys. Rev. B {\bf 70}, 214109 (2004).

\bibitem{Chang2010}
Y. J. Chang, A. Bostwick, Y. S. Kim, K. Horn and E. Rotenberg, Phys. Rev. B {\bf 81}, 235109 (2010).

\end{thebibliography}
\end{document}